\documentclass[runningheads]{svmult}

\usepackage{makeidx}   % allows index generation
\usepackage{graphicx}  % standard LaTeX graphics tool
\usepackage{subeqnar}  % subnumbers individual equations
\usepackage{multicol}  % used for the two-column index
\usepackage{physprbb}  % modified textarea for proceedings,
\makeindex             % used for the subject index
\newcommand{\z}{&&\hspace*{-1cm}}

\newcommand{\vph}{\varphi}
\newcommand{\bea}{\begin{eqnarray}}
\newcommand{\eea}{\end{eqnarray}}
\begin{document}
\title*{The Structure Functions $F_2^c$, $F_L$  and $F_L^c$
\protect\newline in the Framework of the $k_T$ Factorization}
\toctitle{The Structure Functions $F_2^c$, $F_L$  and $F_L^c$
\protect\newline in the Framework of the $k_T$ Factorization}
% allows explicit linebreak for the table of content
%
%
\titlerunning{The Structure Functions $F_2^c$, $F_L$  and $F_L^c$}
% allows abbreviation of title, if the full title is too long
% to fit in the running head
%
\author{
%Anatoly
A.V. Kotikov\inst{1}
\and 
%Artyom 
A.V. Lipatov\inst{2}
\and 
%Gonzalo 
G. Parente\inst{3}
\and 
%Nikolai 
N.P. Zotov\inst{4}}
\authorrunning{A.V. Kotikov et al.}
% if there are more than two authors,
% please abbreviate author list for running head
%
%
\institute{
%Inst. fuer Theor. Teilehenphysik, 
Univ. Karsruhe,
D-76128 Karsruhe, Germany and \\
BLThPh, JINR, 141980 Dubna, Russia
\and Dep. of Physics, MSU, 119992  Moscow, Russia
\and Univ. de Santiago de Compostela,
15706 Santiago de Compostela, Spain
\and SINP, MSU, 119992  Moscow, Russia
}

\maketitle              % typesets the title of the contribution

\begin{abstract}
We present 
the perturbative parts of the structure functions
$F_2^c$ and $F_L^c$ for a gluon target having nonzero transverse
momentum squared at order $\alpha _s$.
The results of the double convolution (with respect to the Bjorken variable
$x$ and the
transverse momentum) of the perturbative part and the unintegrated
gluon densities are compared with HERA experimental data for
$F_2^c$ and  $F_L$ at low $x$ values and with predictions of other 
approaches.
The contribution from $F_L^c$ structure function ranges $10\div30\%$
of that of $F_2^c$ at the HERA kinematical range.
\end{abstract}

\section{Introduction} \indent 

The basic information on the internal
%quark
structure of nucleons is extracted from the process of deep inelastic
(lepton-hadron) scattering (DIS). Its differential cross-section has the form:
\begin{eqnarray}
 \frac{d^2 \sigma}{dxdy}~=~ \frac{2 \pi \alpha_{em}^2}{xQ^4}~~ \bigl[
\left( 1 - y + y^2/2
\right) F_2(x,Q^2) - \left(y^2/2\right) F_L(x,Q^2) \bigr],
%\label{0.1} 
\nonumber \end{eqnarray}
where $F_2(x,Q^2)$ and $F_L(x,Q^2)$ are the transverse and
longitudinal structure functions (SF), respectively,
$q^{\mu}$ and  $p^{\mu}$ are the photon and the hadron 4-momentums and
$x=Q^2/(2pq)$ with $Q^2 = -q^2>0$. 

In the lecture we will study only the charm part $F_2^c$ of the transverse 
SF $F_2$ and the longitudinal SF $F_L$.

The study is related with the fact that recently there have been 
important new data on the charm SF
%structure function (SF) 
$F_2^c$, of the proton from the H1 \cite{H1} and ZEUS \cite{ZEUS} 
Collaborations at HERA, which have probed the small-$x$ region down to 
$x=8\times 10^{-4}$ and $x=2\times 10^{-4}$, respectively. At these values 
of $x$, the charm contribution to the total proton SF, $F_2$, is found
to be around $25\%$, which is a considerably larger fraction than that found 
by the European Muon Collaboration (EMC) at CERN \cite{EMC} at 
%samewhat 
larger $x$, where it was only $\sim 1\%$ of $F_2$. Extensive 
theoretical analyses 
in recent years have generally served to confirm that 
%the bulk of 
the
$F_2^c$ data can be described through perturbative generation of charm 
within QCD (see, for example, the review in Ref. \cite{CoDeRo} and references 
therein).

The second object of our study is
the longitudinal SF $F_L(x,Q^2)$. It 
is a very sensitive QCD characteristic because it is
equal to zero in the parton model with spin$-1/2$ partons.
Unfortunately, essentially at
small values of $x$, the experimental extraction of $F_L$ data 
requires a
rather cumbersome procedure (see \cite{1.5,BaGlKl}, for example).
Moreover, perturbative QCD leads to some controversial results in the 
case of SF $F_L$.
The next-to-leading order (NLO) corrections to the longitudinal 
coefficient function, which are 
large and negative at small $x$ 
\cite{Keller,Rsmallx}, need a resummation procedure 
\footnote{ Without a resummation the NLO
%LO$\&$NLO 
approximation to $F_L$ 
can be negative at 
low $x$ and quite low $Q^2$ values (see \cite{Rsmallx,Thorne02}).}
that leads to a
coupling constant scale higher essentially than $Q^2$ 
(see \cite{DoShi,Rsmallx,Wong})
\footnote{Note that at low $x$
a similar property has been observed
also in 
%BFKL-motivated 
the approaches  \cite{BFKLP,bfklp1,Salam}
(see recent review \cite{Andersson} and discussions therein),
which based on Balitsky-Fadin-Kuraev-Lipatov (BFKL) dynamics
\cite{BFKL}, where the leading $\ln(1/x)$ contributions are summed.}.

Recently there have been important new data \cite{H1FL97}-\cite{Gogi}
of the longitudinal SF
%structure function (SF) 
$F_L$, 
which have probed the small-$x$ region down to $x \sim 10^{-2}$. 
Moreover, the SF $F_L$  can be related at small $x$ with SF $F_2$ 
and the derivative $dF_2/d\ln Q^2$ (see 
%\cite{KoFL}-\cite{KoPaR}
\cite{KoPaFL}). In 
this way most precise predictions based on data of $F_2$ 
and $dF_2/d\ln Q^2$ (see \cite{H1FL} and references therein)
 can be obtained for $F_L$.
% (see Section 4). 
These predictions can be 
considered as indirect 'experimental data' for $F_L$.

We note, 
%however, 
that perhaps more relevant analyses of the HERA data, where 
the $x$ values are quite small, are those based on BFKL dynamics
\cite{BFKL},
% (see discussions in the review of Ref. \cite{Kwien} and references
%therein), 
because the leading $\ln(1/x)$ contributions are summed. The basic
dynamical quantity in the BFKL approach is the unintegrated gluon 
distribution
$\vph_g(x,k^2_{\bot})$ 
\footnote{Hereafter
%$p^{\mu}$ and 
$k^{\mu}$ is
%are the hadron and 
the gluon 
%4-momentums, respectively, and $q^{\mu}$ is the photon 
4-momentum.}
%$\Phi(x_B,k^2_t)$ 
($f_g$ is the (integrated) gluon distribution multiplied
by $x$ and $k_{\bot}$ is the transverse momentum)
 \begin{eqnarray}
f_{g}(x,Q^2) ~=~ f_{g}(x,Q^2_0) + \int^{Q^2}_{Q^2_0} dk^2_{\bot}
%\frac{dk^2_{\bot}}{k^2_{\bot}} 
\; \vph_g(x,k^2_{\bot}), 
\label{1.1}
 \end{eqnarray}
which satisfies the BFKL equation.  
%Notice that the 
The integral is divergent at the lower limit and it leads 
to the necessity to use the difference $f_{g}(x,Q^2) - f_{g}(x,Q^2_0)$
with some nonzero $Q^2_0$.

Then, in the BFKL-like approach (hereafter the $k_t$-factorization
approach \cite{CaCiHa,CoEllis} is used)
the SF $F^c_{2,L}(x,Q^2)$ and $F_{L}(x,Q^2)$  are driven at small 
$x$ by gluons and are related in the following way to the unintegrated 
distribution $\vph_g(x,k^2_{\bot})$: 
\begin{eqnarray}
%\z
F^c_{2,L}(x,Q^2) &=&\int^1_{x} \frac{dz}{z} \int dk^2_{\bot}
%\frac{dk^2_{\bot}}{k^2_{\bot}} 
\, e_c^2
\; C^g_{2,L}(z,Q^2,m_c^2,k^2_{\bot})~ \vph_g(x/z, k^2_{\bot}), 
 \label{d1}\\
%\z 
F_{L}(x,Q^2) &-&
F^c_{L}(x,Q^2) \nonumber \\
&=&\int^1_{x} \frac{dz}{z} \int dk^2_{\bot}
%\frac{dk^2_{\bot}}{k^2_{\bot}} 
%\nonumber \\
%\z ~~~~~ 
\, \sum_{i=u,d,s} e_i^2
\; C^g_{2,L}(z,Q^2,0,k^2_{\bot})~ \vph_g(x/z, k^2_{\bot}), 
 \label{d1a}
%\nonumber
\end{eqnarray}
where $e_i$ is the charge of the $i$-flavor quark.

The functions $C^g_{2,L}(x,Q^2,m_c^2,k^2_{\bot})$ and
$C^g_{L}(x,Q^2,0,k^2_{\bot})$ 
may be regarded as the structure 
functions of
the off-shell gluons with virtuality $k^2_{\bot}$ (hereafter we call them as
{\it hard structure functions} \footnote{This 
notation reflects the fact that SF
%structure functions  
$F_{2,L}^c$ and $F_{L}$ connect with the functions $C_{2,L}^g$
and $C_{L}^g$ at the same form 
as cross-sections connect with hard ones (see \cite{CaCiHa,CoEllis}).}). 
They are described by the quark box (and
crossed box) diagram contribution to the photon-gluon interaction 
(see Fig.1.)
% in \cite{KoLiPaZo}). 

\begin{figure}[t]
\begin{center}
\includegraphics[width=1.0\textwidth]{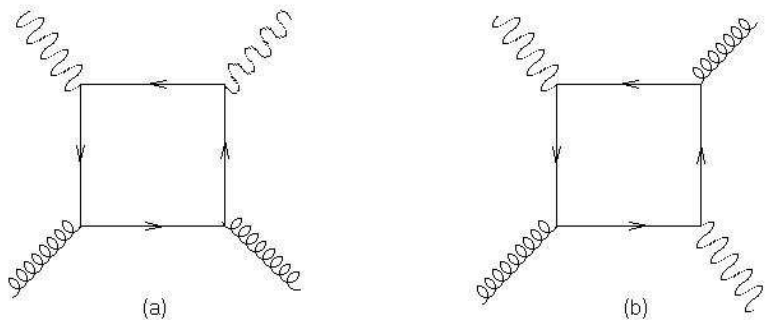}
\end{center}
\caption[]{ The diagrams contributing to $T_{\mu\nu}$ for a gluon target
(the upper and down wave lines on the diagram (a) correspond to photons and 
gluons, respectively).
They should be multiplied by a factor of 2 because of the opposite direction 
of the fermion loop. The diagram (a)
%{\bf (a)} 
should be also doubled because of crossing symmetry.}
%The box diagramms of photon gluon fusion.}
\label{fig1}
\end{figure}

The purpose of the paper is to present the results of \cite{KoLiPaZo} for
%calculate 
these hard SF
$C^g_{2,L}(x,Q^2,m_c^2,k^2_{\bot})$ and to analyze experimental data for 
$F^c_{2}(x,Q^2)$ and $F_{L}(x,Q^2)$
 by applying Eq. (\ref{d1}) with different sets of
unintegrated gluon densities (see Ref. \cite{LiZo,Andersson})
and to give predictions for the longitudinal charm SF $F^c_{L}(x,Q^2)$.
%$F_{L}(x,Q^2)$.\\

It is instructive to note that
the results should be  
similar to those  of the
photon-photon scattering process.
The corresponding QED contributions have been calculated many years ago
in Ref. \cite{BFaKh} (see also the beautiful review in Ref. \cite{BGMS}). 
Our results 
have been calculated independently in \cite{KoLiPaZo} 
(based on approaches of \cite{KaKo,KoTMF})
and they are in full agreement with 
%Ref.
%on 
\cite{BFaKh}. 
However, we hope
that our formulas which are given in a 
%have essentially 
simpler form 
%and we hope that they will 
could be useful for others.
%in other analyses.\\

\section{Hard structure functions} \indent

The gluon polarization tensor (hereafter
the indices $\alpha$ and $\beta$ are connected with gluons),
which gives the main contribution at high energy limit, has the form:
%to the polarization tensor  we are interested in (see below)
\bea
\hat P^{\alpha\beta}_{BFKL}~=~
\frac{k_{\bot}^{\alpha}k_{\bot}^{\beta}}{k_{\bot}^2}
~=~\frac{x^2}{-k^2}
p^{\alpha}p^{\beta}~=~ -\frac{1}{2}
\frac{1}{\tilde \beta^4} \left[\tilde \beta^2 g^{\alpha \beta} 
-12 bx^2 \frac{q^{\alpha}q^{\beta}}{Q^2} \right],
%(k=xp+k_t)
\label{1dd}
\eea
where $\tilde \beta^2=1-4bx^2,~~b=-k^2/Q^2 \equiv  k_{\bot}^2/Q^2 >0.$

Contracting the corresponding photon projectors, we have: 
 \begin{eqnarray}
\z C^g_{2}(x) = 
%e_c^2 \; \frac{\alpha_s(Q^2)}{4\pi}\; x 
%\frac{{\cal K}}{\tilde \beta^2 }
\frac{a_s}{\tilde \beta^2 }
\; x
\;
\left[
f^{(1)} + 
\frac{3}{2\tilde \beta^2}\; f^{(2)} \right],~
% \label{3}\\
C^g_{L} (x) = 
%e_c^2 \; \frac{\alpha_s(Q^2)}{4\pi}\; x 
\frac{a_s}{\tilde \beta^2 }
\;x \,
\left[
4bx^2 f^{(1)} + 
\frac{(1+2bx^2)}{\tilde \beta^2}\; f^{(2)} \right]
%\nonumber 
\label{3}\\
\z f^{(1)} = 
\frac{1}{\tilde\beta^4}
\left[ \tilde \beta^2 \hat f^{(1)}  ~-~
3bx^2 \tilde f^{(1)}\right],~~~
f^{(2)} ~=~
\frac{1}{\tilde\beta^4}
\left[ \tilde \beta^2 \hat f^{(2)}  ~-~
3bx^2 \tilde f^{(2)} \right],
%&=& 
\nonumber
 \end{eqnarray}
where the 
%normalization factor ${\cal K} =  
%%e_c^2 \; 
$a_s(Q^2)=\alpha_s(Q^2)/(4\pi)$, $a=m^2/Q^2$
%, ~$a=m^2_c/Q^2$~, $e_c=2/3$ is charm quark charge 
and
\begin{eqnarray}
\hat f^{(1)} &=& -2 \beta \Biggl[ 1 - \biggl(1-2x(1+b-2a) \; [1-x(1+b+2a)] 
\biggr) \; f_1  
\nonumber \\
    &+& (2a-b)(1-2a)x^2 \; f_2  \Biggr],
 \label{5}\\
\hat f^{(2)} &=& 8x\; \beta \Biggl[(1-(1+b)x)  
-2x \biggl(bx(1-(1+b)x)(1+b-2a) + a\tilde \beta^2 \biggr)\; f_1  
\nonumber \\
    &+& bx^2(1-(1+b)x) (2a-b) \; f_2  \Biggr],
\label{6} \\
\tilde f^{(1)} &=& - \beta \Biggl[ \frac{1-x(1+b)}{x}  
-2 \biggl(x(1-x(1+b))(1+b-2a) +a \tilde \beta^2 \biggr) \; f_1  
\nonumber \\
    &-& x(1-x(1+b))(1-2a) \; f_2  \Biggr],
 \label{5dd}\\
\tilde f^{(2)} &=& 4 \; \beta ~(1-(1+b)x)^2 \Biggl[2   
- (1+2bx^2)\; f_1  
%\nonumber \\    &-& 
-bx^2 \; f_2  \Biggr],
\label{6dd} 
\end{eqnarray}
with
$$
%~~~~~~
\beta^2=1-\frac{4ax}{(1-(1+b)x)},~~~~~f_1=\frac{1}{\tilde \beta \beta} \;
\ln\frac{1+\beta \tilde \beta}{1-\beta \tilde \beta},
~~~~~f_2=\frac{-4}{1-\beta^2 \tilde \beta^2}$$

For the important regimes when
$k^2= 0$, $m^2_c = 0$ and/or $Q^2 = 0$, the 
%analyses are given in  Appendix C.
results coincide with ones in Refs. \cite{CaCiHa,Witten}.
Notice that our results in Eq. (\ref{3})  should  also
agree with those in Ref. \cite{BMSS} but the direct comparison is 
quite difficult because 
the structure of their results is quite cumbersome (see Appendix A in 
Ref. \cite{BMSS}). 
%\\
We have found numerical agreement in the case of 
$F_2(x,Q^2)$ for several types of unintegrated gluon distributions
(see Fig.4 in \cite{KoLiPaZo}).

%%%%%%%%%%%%%% 3  3  3
\section{Relations between $F_L$, $F_2$ and derivation of $F_2$
in the collinear approximation} \indent

%The 
Another information about the SF $F_L$ can be obtained in the 
%framework of 
collinear approximation (i.e. when $k_{\bot}^2 =0$) in the following way
(see also our study \cite{KoLiZo}).

In the framework of perturbative QCD, there is the possibility to connect
$F_L$ to $F_2$  and $dF_2/d\ln Q^2$ due the fact that at 
small $x$ the DIS structure functions 
depend only on two {\it independent} functions: the gluon distribution and
the singlet quark one (the nonsinglet quark density is negligible at 
small $x$),
which in turn can be expressed in terms of measurable $F_2$ and 
$dF_2/d\ln Q^2$.

In this way, by analogy with the case of the gluon distribution function (see
\cite{Prytz,KoPaG} and references therein), the behavior of
$F_L(x,Q^2)$ has been studied in \cite{KoPaFL},
%\cite{KoFL}-\cite{KoPaR},
% atsmall values of $x$, 
using the 
%quite old 
HERA data \cite{F2H1,F2ZEUS} and the method \cite{method1}
\footnote{The method is based on previous investigations \cite{1.5,method}.}
of replacement of the Mellin convolution by ordinary products.
Thus, the small $x$ behavior of
$F_L(x,Q^2)$ can be extracted directly from the measured values
of $F_2(x,Q^2)$ and its derivative without a cumbersome procedure (see
\cite{1.5,BaGlKl}).  These extracted values of $F_L$ may be well considered as
{\it new
small $x$ 'experimental data' of $F_L$}.
The relations can be violated by nonperturbative corrections like 
higher twists (see \cite{HT,KriKo}), which can be large 
in the case of $F_L$ \cite{CaHa,BaGoPe}. 

Because $k_T$-factorization approach is one of popular perturbative
approaches used at small $x$, it is very useful to compare its predictions
with the results of \cite{KoPaFL}
%-\cite{KoPaR} 
based on
%lead to 
the relations
between SF $F_L(x,Q^2)$, $F_2(x,Q^2)$
and $dF_2(x,Q^2)/d\ln Q^2$. 
%It is the main purpose of the study.

%However, the 
The $k_T$-factorization approach is strongly connected to the Regge-like 
behavior of 
parton distributions.
So, we restrict our investigations to 
%a Regge-likeform of 
SF
%structure functions 
and parton distributions with the following form (hereafter $a=q,g$):
%, one obtains (see \cite{KoMPL})
\begin{eqnarray} f_a(x,Q^2) \sim F_2(x,Q^2) \sim  x^{-\delta(Q^2)} 
\label{1.9} \end{eqnarray}

Note that really the slopes of the sea quark and gluon distributions:
$\delta_q$ and $\delta_g$, respectively, and the slope $\delta_{F2}$
of $F_2$ are sligtly different.
The slopes have a familiar property $\delta_q <\delta_{F2} 
<\delta_g $ (see Refs. \cite{H1slo}-\cite{KoMPL} and references therein).
We will neglect, however, this difference and use
%Moreover, in %In 
in our investigations the experimental values of 
$\delta(Q^2) \equiv \delta_{F2}(Q^2)$ 
extracted by H1 
%and ZEUS 
Collaboration 
\footnote{ Now the preliminary ZEUS data for the slope $d\ln F_2/d\ln (1/x)$
are available as some points on Figs. 8 and 9 in Ref. \cite{slope}.
Moreover, the new preliminary H1 points have been presented on the 
Workshop DIS2002 (see \cite{DIS02}). Both the new points
%They 
are shown quite similar properties to compare with H1 data 
\cite{H1slo}. Unfortunately, tables of the ZEUS data and the new H1 data are
 unavailable yet
and, so, the points cannot be used here.} (see \cite{H1slo}
and references therein).
We 
%would like to 
note that the $Q^2$-dependence is in very good agreement
with perturbative QCD at $Q^2 \geq 2$ GeV$^2$ (see \cite{KoPaslo}). 
Moreover,
the values of the slope $\delta(Q^2)$ are 
in agreement with recent phenomenological studies (see, for example,
\cite{BFKLP}) incorporating the NLO 
%next-to-leading 
corrections \cite{FaLi} (see also \cite{KoLi})
in the framework of BFKL approach.

 Thus, assuming the {\it Regge-like behavior} (\ref{1.9})
for the gluon distribution and
 $F_2(x,Q^2)$ at $x^{-\delta} \gg 1$
and using the NLO 
approximation for collinear coefficient functions and anomalous
dimensions of Wilson operators,
%of $r^{1+\delta}_{sp}$ and $r^{1+\delta}_{Lp}$
the following approximate
results for F$_L(x,Q^2)$ has been obtained in \cite{KoLiZo}:

\begin{eqnarray}   
\z  F_L(x, Q^2)  = 
%\frac{4(1+ \delta)}{(4+ 3\delta + \delta^2)}
\frac{r(1+\delta){(\xi(\delta))}^{\delta}}{ (1 + 30 a_s(Q^2) [
1/\tilde \delta - \frac{116}{45} \rho_1(\delta)])  } \biggl[
\frac{d F_2(x\xi(\delta), Q^2)}{d\ln Q^2} 
\nonumber \\ \z
~~~~~~+~ \frac{8}{3} ~
\rho_2(\delta) a_s(Q^2)  F_2(x\xi(\delta), Q^2)   \biggr] 
~+~ 
O(a_s^2, a_s x, x^2), 
\label{10.51} 
\end{eqnarray} 
where
\begin{eqnarray}
%\frac{1}{\delta} \stackrel{\delta \to 0}{\to} 
r(\delta) &=& \frac{4\delta}{2+\delta+\delta^2},~~~
\xi(\delta) ~=~ \frac{r(\delta)}{r(1+\delta)},\nonumber \\
\rho_1(\delta) &=& 1+ \delta + \delta^2/4,~~~
\rho_2(\delta) ~=~ 1-2.39 \delta + 2.69 \delta^2, 
\label{10.6} \end{eqnarray}

Here
\begin{eqnarray}
%\frac{1}{\delta} \stackrel{\delta \to 0}{\to} 
\frac{1}{\tilde \delta}
&=& \frac{1}{\delta}
\biggl[
1 ~-~ \frac{\Gamma(1- \delta)\Gamma(1+ \nu)}{\Gamma(1- \delta + \nu)} 
x^{\delta} \biggr],
\label{d1.2} \end{eqnarray}
where $\Gamma$ is the Rimmanian $\Gamma$-function.
The value of $\nu$ comes (see \cite{TMF})
from asymptotics  of parton distributions 
$f_a(x)$ at $x \to 1$: $f_a \sim (1-x)^{\nu_a}$,
and 
\footnote{In our 
%small $x$ 
formula (\ref{d1.2}) we are mostly interested
in gluons, so we take $\nu =\nu_g \approx 4$ below.}
$\nu \approx 4$ from quark account rules \cite{schot}.

%We would like to note 
Note that the $1/\tilde \delta$
coincides approximately with $1/\delta$ when $\delta \neq 0$ and 
$x \to 0$. However, at $\delta \to 0$, the value of $1/\tilde \delta$ is not 
singular:
\begin{eqnarray}
\frac{1}{\tilde \delta} 
\to 
\ln \left(\frac{1}{x}\right) - \biggl[ \Psi(1+ \nu) - \Psi(1) \biggr], 
\label{10.7} \end{eqnarray}
where
$\Psi$-function is the logarithmic derivation of the $\Gamma$-function.

%Note also that the results (\ref{10.51}) are exact at 
%$\delta = 0, 0.3$ and $0.5$
%values (see \cite{KoLiZo} and duscussions therein).

%%%%%% 4

\section{Comparison with $F_2^c$ and $F_L$ experimental data and 
predictions for $F_L^c$ } \indent

%\vskip 0.5cm
With the help of the results obtained in the previous sections
we have analyzed HERA data \cite{ZEUS} for SF $F_2^c$ from ZEUS   
Collaboration and the data for SF $F_L$ mostly from H1 
%\cite{H1} 
Collaboration.

Notice that in Ref. \cite{CaCiHa} 
the $k_{\bot}^2$-integral in the r.h.s. of Eq.(\ref{d1}) has been evaluated
using the BFKL results for the Mellin transform of the
unintegrated gluon distribution and the Wilson coefficient functions have 
been calculated for the full perturbative series at asymptotically small
$x$ values. Since we would like to analyze experimental data for 
$F_2^c$ and $F_L$ 
%we have an interest to obtain results at 
on a quite broad range of 
small $x$ values,
% For the reason 
we need a parameterization
of unintegrated gluon distribution.

To study $F_2^c$ we consider two different parametrizations for the
unintegrated gluon distribution (see ~\cite{LiZo}):
%. Firstly,  we use the parametrization 
the Ryskin-Shabelski (RS) one ~\cite{RS}
and the Blumlein one 
\cite{Blumlein}. 
\begin{figure}
%
%\vskip -1cm
\begin{center}
%\vskip -1cm
%\epsfig{figure=f2ccZEUSx.eps,width=15cm,height=20cm}
%\epsfig{figure=f2ccZEUSx.eps,width=16cm,height=21.2cm}
%\epsfig{figure=f2ccZEUSx.eps,width=16.5cm,height=21.8cm}
%\includegraphics[width=.3\textwidth]{f2ccZEUSx.eps}
\includegraphics[width=1.0\textwidth]{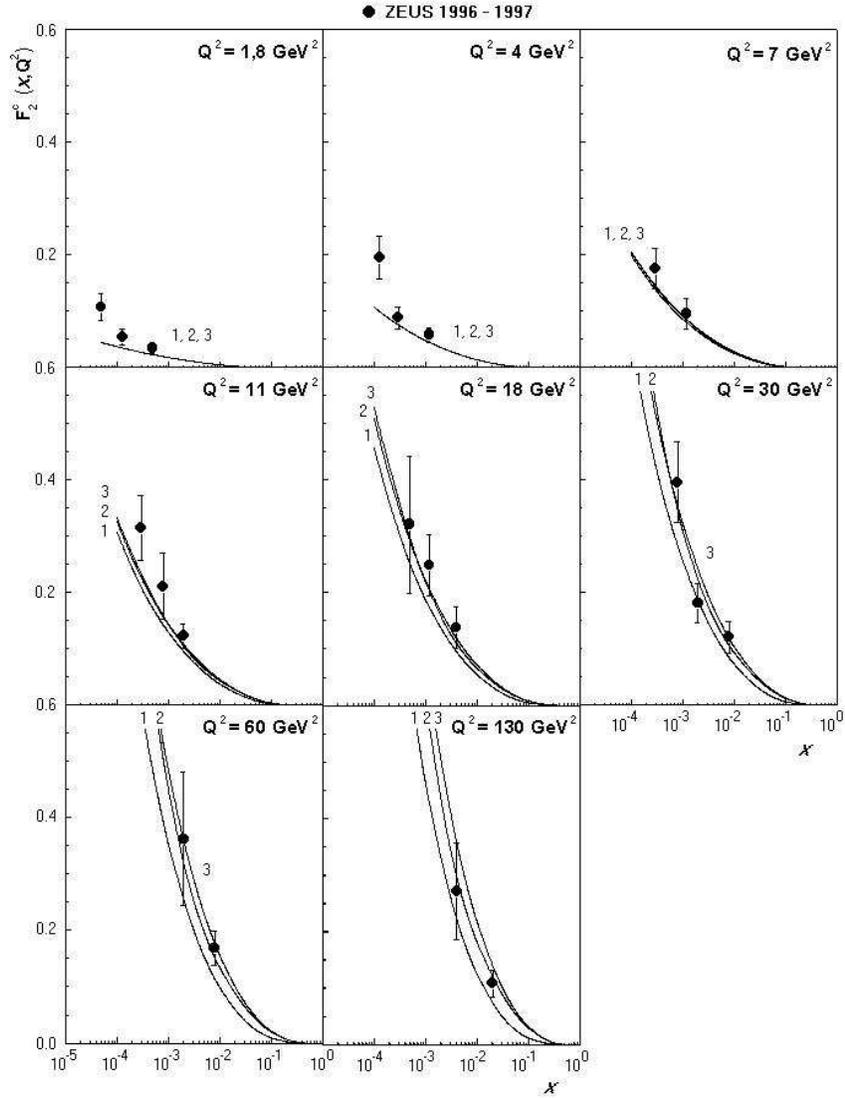}
\end{center}
%\vskip -1cm
\caption[]{The structure function $F_2^c(x,Q^2)$ as a function of $x$ for
different values of $Q^2$ compared to ZEUS data~\cite{ZEUS}.
Curves 1, 2 and 3 correspond, respectively, to 
%SF obtained in 
pure perturbative QCD
%the standard parton model 
with the GRV ~\cite{GRV}
gluon density at the leading order approximation and 
%to SF obtained in the 
$k_T$ factorization 
%approach 
with RS~\cite{RS} 
and Blumlein (at $Q_0^2 = 4$ GeV$^2$)~\cite{Blumlein} parameterizations of 
unintegrated gluon distribution.}
\label{fig2}
\end{figure}

%In Fig. 2 
%%and 2 
%we show  the SF $F_2^c$ as a function $x$ for different values
%of $Q^2$  in comparison with ZEUS \cite{ZEUS} 
%%and H1 \cite{H1}
%experimental data.
%
 We see in Fig. 2 that at large $Q^2$ ($Q^2 \geq 10$ GeV$^2$)
the SF $F_2^c$
obtained in the $k_T$ factorization approach is higher than the SF obtained
in pure perturbative QCD
%the standard parton model 
with the GRV \cite{GRV}
%and MT ~\cite{MT} 
gluon density at the LO approximation 
(see curve 1) 
and has a more rapid growth in comparison with perturbative QCD
%the standard parton model
results, especially at $Q^2 \sim 130$ GeV$^2$~\cite{LZ}.
For $Q^2 \leq 10$ GeV$^2$ the predictions from
perturbative QCD 
%(in GRV approach)
and those based on the $k_T$ factorization approach are very similar
\footnote{This fact is also due to the quite large value of $Q^2_0=4$ 
GeV$^2$ chosen here.} and show a disagreement with data 
below $Q^2 = 7$ GeV$^2$
\footnote{A similar disagreement with data at $Q^2 \leq 2$ GeV$^2$
has been observed for the complete structure function $F_2$
(see, for example, the discussion in Ref. \cite{Q2evo}
and reference therein). 
We note that the insertion of higher-twist corrections in the framework
of usual perturbative QCD improves the agreement with data
(see Ref. \cite{HT}) at quite low values of $Q^2$.}.
Unfortunately the available experimental data do not permit yet
to distinguish 
%so far 
the $k_T$ factorization effects from
those due to boundary conditions~\cite{RS}.

The 
%difference between the 
results for the SF  $F_L^c$ obtained in perturbative QCD and 
from the $k_T$ factorization approach are quite similar to the $F_2^c$ case
discussed above.
The ratio $ R^c = F_L^c/F_2^c$ is 
shown in Fig. 3. We see that $ R^c \approx 0.1 \div 0.3 $ in a wide 
region of $Q^2$.
%The estimation of $R^c$ is very close to the results for $R=F_L/(F_2-F_L)$
%ratio (see Refs. \cite{KoPaFL}-\cite{CCFRr}).
We would like to note that these values of $ R^c $
contradict the estimation
obtained in Ref.~\cite{ZEUS}. So, the effect of the large $ R^c $ values
should be considered in the extraction of $F_2^c$
from the corresponding differential cross-section
in future more precise measurements.

\begin{figure}
%\begin{center}
%\vskip -1cm
%\epsfig{figure=RccZEUSx.eps,width=17cm,height=20cm}
%\epsfig{figure=RccZEUSx.eps,width=17.5cm,height=20.8cm}
%\includegraphics[width=.3\textwidth]{RccZEUSx.eps}
\includegraphics[width=1.0\textwidth]{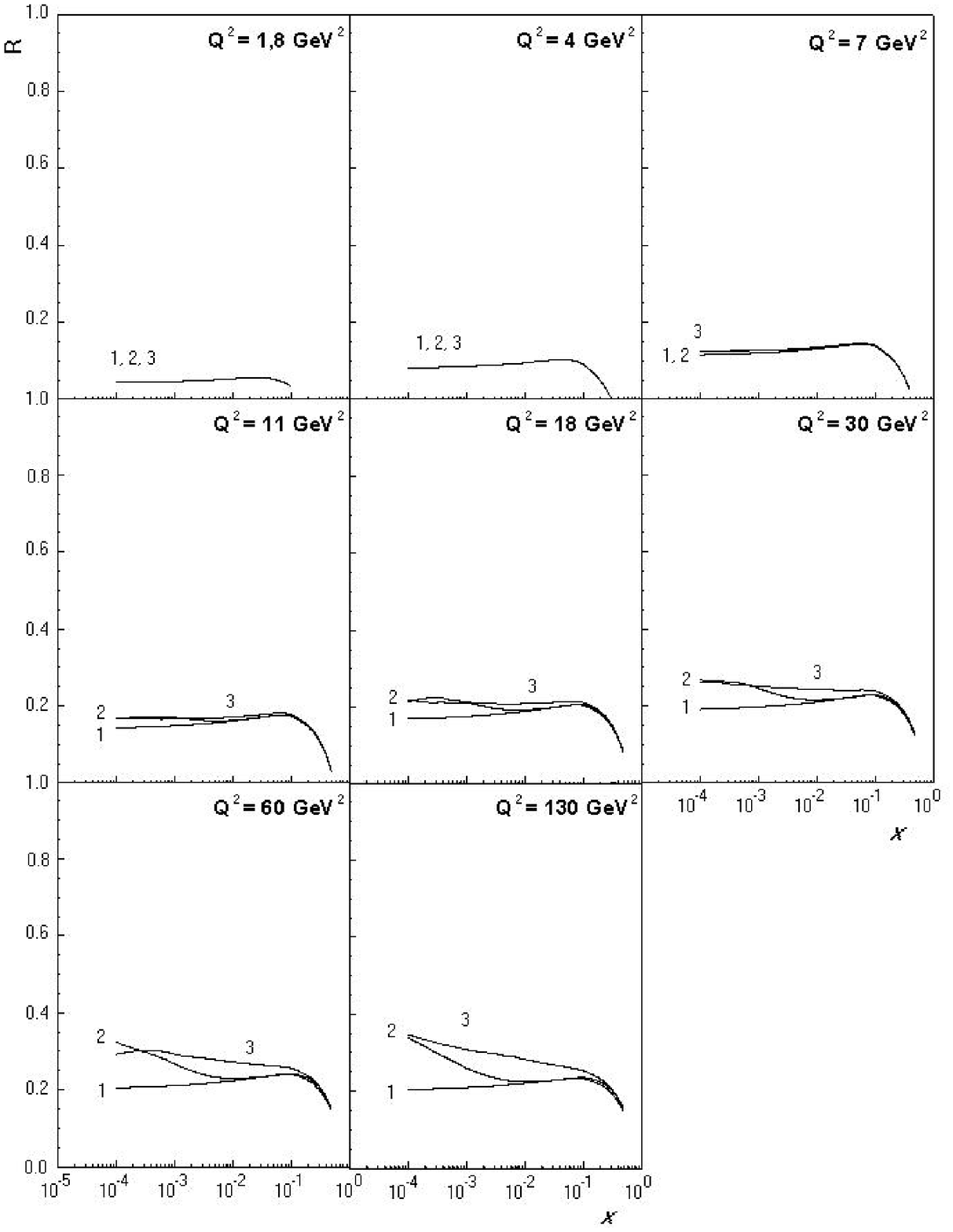}
%\end{center}
%\vskip -0.5cm
\caption[]{The ratio  $R^c  = F_L^c(x,Q^2)/F_2^c$ as a function of $x$ 
for different values of $Q^2$.
Curves 1, 2 and 3 are as in Fig. 2.}
\label{fig3}
\end{figure}

For the ratio $ R^c $ we found quite flat $x$-behavior at low $x$ 
in the low $Q^2$ region (see Fig. 3), where approaches based on perturbative
QCD and on $k_T$ factorization give similar predictions 
(see Fig 2).
%$Q^2 \leq 60$ GeV$^2$.
It is in agreement with the corresponding behaviour of the ratio
$R=F_L/(F_2-F_L)$ (see Ref. \cite{KoPaFL}) at quite large values of $\delta $
\footnote{The behaviour is in agreement with previous studies 
\cite{TMF,KoPaFL}. Note that
at small values of $\delta $, i.e. when $x^{-\delta} \sim Const$,
%the ratio 
$R$ has the strong negative NLO corrections 
(see Refs. \cite{Keller,Rsmallx})
and tends to zero at $x \to 0$ after some resummation done in \cite{Rsmallx}.}
($\delta > 0.2-0.3$).
The low $x$ rise of $ R^c $ at high $Q^2$ disagrees with early
calculations \cite{KoPaFL} in the framework of perturbative QCD.
It could be due to the small $x$ resummation, which is important at high $Q^2$
(see Fig 2).
We plan to study this effect in future.
% on $R$ .

In Fig. 4 we show  the SF $F_L$ as a function $x$ for different values
of $Q^2$  in comparison with H1 experimental sets: the old one 
\cite{H1FL97} 
(black triangles), the last year set \cite{H1FL} (black squares) and 
the new preliminary data \cite{Gogi} (black circles) and also with NMC 
\cite{NMC} (white triangles), CCFR \cite{CCFR} (white circles) and BCDMS
\cite{BCDMS} data (white squares).
For comparison with these data
we present the results of the calculation with three 
different 
parameterizations for the unintegrated gluon distribution 
$\vph_g (x, k^2_{\bot}, Q_0^2)$ 
%in the forms given by Eq. (\ref{zo1})\, and Eq. (\ref{zo1dd}) 
at $Q_0^2 = 4$ GeV$^2$: 
%All of them: 
%Ryskin-Shabelsky (RS) 
Kwiecinski-Martin-Stasto (KMS)  \cite{KMS},
%from \cite{RySha} and 
Blumlein \cite{Blumlein} and Golec-Biernat and Wusthoff (GBW)
 \cite{GBW}. 
%have been used already in our 
%previous work \cite{KLPZ} and reviewed there.
% (see also \cite{LiZo}).
%The third unintegrated gluon function used here is one proposed by
%Golec-Biernat and Wusthoff (GBW) which takes into account
%saturation effects and has been applied earlier in analysis of the 
%inclusive and diffractive $ep$-scattering data \cite{GoBiWu}. 
%%It hast the following form:\\

%Fig. 5 is similar to Fig. 4 with one excepting: 
We also add  the 'experimental data' obtained using the relation 
between SF $F_L(x,Q^2)$, $F_2(x,Q^2)$
and $dF_2(x,Q^2)/d\, {\ln Q^2}$ (see Section 3) as black stars. Because 
the corresponding data for 
$F_2(x,Q^2)$ and $dF_2(x,Q^2)/d\, {\ln Q^2}$ are essentially more precise
(see \cite{H1FL})
to compare with the preliminary data \cite{Gogi} for $F_L$, 
%the uncertainties of 
the 'experimental data' have strongly suppressed uncertainties.
%are essentially less.
As it is shown in Fig. 4 there is very good agreement between the new
preliminary data \cite{Gogi}, the 'experimental data' and predictions
from perturbative QCD and $k_T$-factorization approach.

 The differences observed between the curves 2, 3 and 4 are 
 due to the different behavior of the unintegrated gluon distribution
as a function of $x$ and $k_{\bot}$.
 We see that 
%at large $Q^2$ ($Q^2 \geq 7$ GeV$^2$)
the SF $F_L$
obtained in the $k_T$-factorization approach with KMS and 
Blumlein parameterizations
is close each other 
%\footnote{Note that very similar results
%have been obtained also for RS
%%Ryskin-Shabelsky 
%parameterization \cite{RS} 
%(see \cite{KoLiZo}).}
and higher than the SF obtained
in the pure perturbative QCD
%standard parton model 
with the GRV 
%and MT ~\cite{MT} 
gluon density at the leading order approximation.
Otherwise, the $k_T$-factorization approach with GBW parameterization
is very close to pure QCD predictions:
%\footnote{This 
%fact is evident also from quite large value of $Q^2_0=4$ 
%GeV$^2$ chosen here.}: 
it should be so because 
GBW model has deviations from perturbative QCD only at quite low $Q^2$
values.
Thus, the predictions from perturbative QCD 
%(in GRV approach)
and those based on $k_T$ factorization approach are in
%quite similar and show a good 
agreement with each other and with all data 
within modern experimental uncertainties. So, a possible high values
of high-twist corrections to SF $F_L$ predicted in \cite{BaGoPe}
can be important only at 
%quite
low $Q^2$ values: $Q^2 \leq Q^2_0=4$ GeV$^2$.

\begin{figure}
\begin{center}
\includegraphics[width=1.0\textwidth]{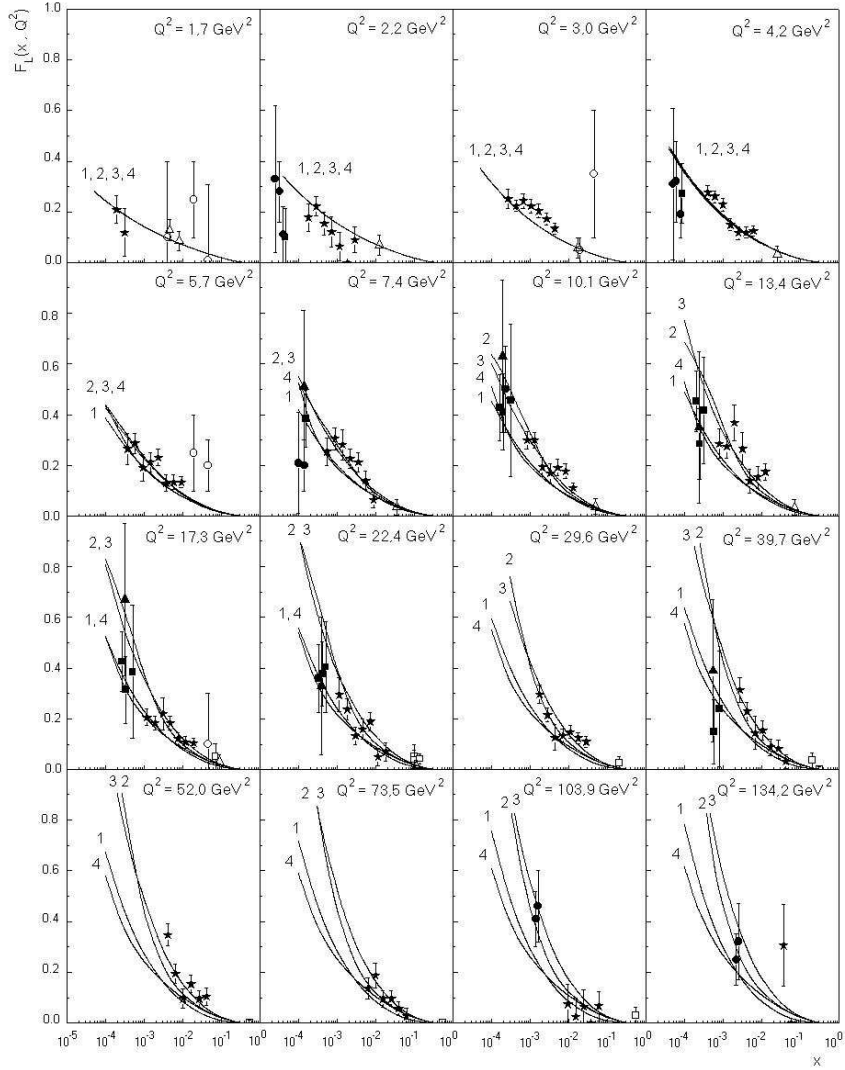}
\end{center}
\caption{The structure function $F_L(x,Q^2)$ as a function of $x$ for
different values of $Q^2$ compared to experimental data. The H1 data:
the first 1997 ones ~\cite{H1FL97},
the new 2001 set ~\cite{H1FL} and the preliminary data ~\cite{Gogi}
are shown as black triangles, squares and circles, respectively.
The NM \cite{NMC}, CCFR \cite{CCFR} and BCDMS \cite{BCDMS} data
are shown as white triangles, circles and squares, respectively.
The 'experimental data'
are added as black stars.
Curves 1, 2, 3 and 4 correspond to SF obtained
in the perturbative QCD with the GRV ~\cite{GRV}
gluon density at the leading order approximation and to SF
obtained in the $k_T$ factorization approach with KMS~\cite{KMS}, 
Blumlein (at $Q_0^2 = 4$ GeV$^2$)~\cite{Blumlein} and GBW \cite{GBW} 
parametrizations of unintegrated gluon distribution.
}
\label{fig4}
\end{figure}
%

%Note that 
There are several other popular parameterizations (see, 
for example,
Kimber-Martin-Ryskin (KMR)  \cite{Ryskin} and Jung-Salam (JS)  
\cite{JuSa}),
which are not used in our study mostly because of technical difficulties.
%\footnote{Note that the RS parameterization \cite{RS} is quite old. 
%We use it together with the JB set \cite{BL} (when the value 
%$\mu^2 = Q_0^2 = 1$ GeV$^2$) only one time (see Fig. 6) to prove
%the coincidence between our off mass shell matrix elements and those from 
%Ref.~\cite{BMSS}.}. 
Note that all above parameterizations give quite similar results  excepting, 
perhaps, the contributions from the small $k_{\bot}^2$-range: 
$k_{\bot}^2 \leq 1$ GeV$^2$ 
(see Ref. \cite{Andersson} and references therein). Because we use 
$Q_{0}^2 =4$ GeV$^2$
in the study of SF $F_{2}^c$ and $F_{L}$, our results depend very slightly  
on the the small $k_{\bot}^2$-range of the parameterizations.
In the case RS, Blumlein, GBW and KMS sets this observation is supported
below by our results and we expect that the application of KMR and JS sets
should not strongly change our results.

%\vspace{-0.2cm}
\section{Conclusions} \indent
%\vspace{-0.5cm}

We have presented
%performed 
the results for
%calculation of 
the perturbative
parts of the SF
%structure functions 
$F_2^c$ and $F_L^c$ for a gluon target 
having nonzero momentum squared, in the process 
of photon-gluon fusion.

We have applied the
results in the framework of $k_T$ factorization approach
to the analysis of
present data for the SF $F_L$ and for the charm contribution to $F_2$ 
%($F_2^c$)
and we have given the predictions for $F_L^c$.
The analysis has been performed with
several parameterizations of unintegrated gluon
distributions,  for comparison. 

For SF $F_2^c$,  we have found good agreement
of our results
%, obtained with RS and BFKL parametrizations of
%unintegrated gluons distributions at $Q_0^2 = 4$ GeV$^2$,
with experimental HERA data,
% for $F_2^c$,
except at low $Q^2$ ($Q^2 \leq 7$ GeV$^2$).
%\footnote{It must be noted that the cross section of inelastic $c\bar c-$
%and $b\bar b-$pair photoproduction at HERA are described by the
%Blumlein parametrization at a smaller value of $Q_0^2$ ($Q_0^2 = 1$ GeV$^2$)
%(see \cite{LiZo}).}. 
We have  obtained also
quite large contribution of the SF $F_L^c$ at low $x$  and
high $Q^2$ ($Q^2 \geq 30$ GeV$^2$) and
this effect 
%of $ R^c $
should be considered for the extraction of $F_2^c$
from the corresponding differential cross-section
in future more precise measurements.

We would like to note the good agreement between our results for $F_2^c$
and the ones obtained in Ref. \cite{Jung} by Monte-Carlo studies. Moreover, we
have also good agreement with fits of H1 and ZEUS data for $F_2^c$
(see recent reviews in Ref. \cite{Wolf} and references therein)
based on perturbative
QCD calculations. But unlike to these fits,
our analysis uses universal unintegrated gluon distribution, which gives
in the simplest way the main contribution to the cross-section in the
high-energy limit.

%We have applied in the framework of $k_T$-factorization approach the
%%The application of our 
%results of the calculation of the perturbative
%parts for the structure functions $F_L$ and $F_L^c$ 
%for a gluon target 
%having nonzero momentum square, in the process 
%of photon-gluon fusion
%to the analysis of
%present data for the structure function $F_L$
%\footnote{In Ref \cite{KLPZ} we have also obtained
%quite large contribution of SF $F_L^c$ at low $x$  and
%high $Q^2$ ($Q^2 \geq 30$ GeV$^2$).}.
%%have been given.
%The analysis has been performed with 
%several parameterizations of unintegrated gluon
%distributions, for comparison. 

For SF $F_L$,
we have found good agreement
%, perhaps, except at low $Q^2$ ($Q^2 \leq 7$ GeV$^2$), 
between all existing experimental data, 
%the preliminary data \cite{Gogi}, 
the predictions for 
$F_L$ obtained from the relation between SF $F_L(x,Q^2)$, $F_2(x,Q^2)$
and $dF_2(x,Q^2)/d\ln Q^2$
and the results 
obtained in the framework of 
perturbative QCD 
%(based on GRV approach)
and ones based on $k_T$-factorization approach
 with the three different 
%RS(?) and BFKL(?) 
parameterizations of unintegrated gluon
distributions.

It could be also very useful to evaluate the complete $F_2$ itself and 
the derivatives of $F_2$ with respect to the 
logarithms of $1/x$ and $Q^2$ with our expressions using the unintegrated
gluons.
%We are considering to present this work and also the predictions for
%$F_L$ in a forthcoming article.

The consideration of the SF $F_2$ in the framework of the leading-twist
approximation of perturbative QCD (i.e. for ``pure'' perturbative QCD)
leads to very good agreement (see Ref. \cite{Q2evo} and references therein) 
with HERA
data at low $x$ and $Q^2 \geq 1.5$ GeV$^2$. The agreement improves
at lower $Q^2$ when higher twist terms are 
taken into account \cite{HT}. As
it has been studied in Refs. \cite{Q2evo,HT}, the SF $F_2$  at low 
$Q^2$ is sensitive to the
small-$x$ behavior of quark distributions.
Thus, our future 
analysis of $F_2$ in a broader $Q^2$ range in the framework
of $k_T$ factorization 
should require the incorporation of parametrizations for unintegrated
quark densities, introduced recently (see Ref. \cite{Ryskin} and references
therein).

%
%
%\vspace{1cm}
%\hspace{1cm} \Large{} {\bf Acknowledgments}    \vspace{0.5cm}

%\normalsize{}
\vskip 0.2cm
\large{{\bf Acknowledgments.}}~~

\normalsize{}
Authors
 would like to express their sincerely thanks to the Organizing
  Committee of the International School ``Heavy Quark Physics'' 
for the kind invitation,  
the financial support
% at  such remarkable Conferences, 
and  for fruitful discussions.
%The study is supported in part by the RFBR grant 02-02-17513.
We are grateful also to 
Professor Catani for useful discussions and comments.

%One of the authors (A.V.K.) 
The study is supported in part by the RFBR grant 02-02-17513.
A.V.K. was supported in part
by INTAS  grant N366 and Alexander von Humboldt fellowship
at the beginning and the end of the study, respectively.
G.P. acknowledges the support of Galician research funds
(PGIDT00 PX20615PR) and Spanish CICYT (FPA2002-01161).
N.P.Z. also acknowledge the support of Royal Swedish Academy of
Sciences.

%INDEX%%%%%%%%%%%%%%%%%%%%%%%%%%%%%%%%%%%%%%%%%%%%%%%%%%%%%%%%%%%%%%%
% Please check with the editor of your book whether he plans to
% include a "mutual" subject index - if so, please code your entries
% in the standard syntax. For your own purposes you may print your
% "personal" index by using the following commands:
%
%\clearpage
%\addcontentsline{toc}{section}{Index}
%\flushbottom
%\printindex
%%%%%%%%%%%%%%%%%%%%%%%%%%%%%%%%%%%%%%%%%%%%%%%%%%%%%%%%%%%%%%%%%%%%%

\end{document}